# MicroBundlePillarTrack: A Python package for automated segmentation, tracking, and analysis of pillar deflection in cardiac microbundles

Hiba Kobeissi[1], Xining Gao[2,3,4], Samuel J. DePalma[5], Jourdan K. Ewoldt[2], Miranda C. Wang[2,3,4], Shoshana L. Das[2,3,4], Javiera Jilberto[5], David Nordsletten[5,6,7], Brendon M. Baker[5], Christopher S. Chen[2,4], Emma Lejeune[1,*]

[1]Department of Mechanical Engineering, Center for Multiscale and Translational Mechanobiology, Boston University, Boston, Massachusetts, United States
[2]Department of Biomedical Engineering, Boston University, Boston, Massachusetts, United States
[3]Institute for Medical Engineering and Science, Harvard–MIT Division of Health Sciences and Technology, Cambridge, Massachusetts, United States
[4]Wyss Institute for Biologically Inspired Engineering, Boston, Massachusetts, United States
[5]Department of Biomedical Engineering, University of Michigan–Ann Arbor, Ann Arbor, Michigan, United States
[6]Department of Cardiac Surgery, University of Michigan–Ann Arbor, Ann Arbor, Michigan, United States
[7]School of Imaging Sciences and Biomedical Engineering, King's Health Partners, King's College London, London, England, United Kingdom
[*]Corresponding author: elejeune@bu.edu

## Abstract

Movies of human induced pluripotent stem cell (hiPSC)-derived engineered cardiac tissue (microbundles) contain abundant information about structural and functional maturity. However, extracting these data in a reproducible and high-throughput manner remains a major challenge. Furthermore, it is not straightforward to make direct quantitative comparisons across the multiple in vitro experimental platforms employed to fabricate these tissues. Here, we present "MicroBundlePillarTrack," an open-source optical flow-based package developed in Python to track the deflection of pillars in cardiac microbundles grown on experimental platforms with two different pillar designs ("Type 1" and "Type 2" design). Our software is able to automatically segment the pillars, track their displacements, and output time-dependent metrics for contractility analysis, including beating amplitude and rate, contractile force, and tissue stress. Because this software is fully automated, it will allow for both faster and more reproducible analyses of larger datasets and it will enable more reliable cross-platform comparisons as compared to existing approaches that require manual steps and are tailored to a specific experimental platform. To complement this open-source software, we share a dataset of 1,540 brightfield example movies on which we have tested our software. Through sharing this data and software, our goal is to directly enable quantitative comparisons across labs, and facilitate future collective progress via the biomedical engineering open-source data and software ecosystem.

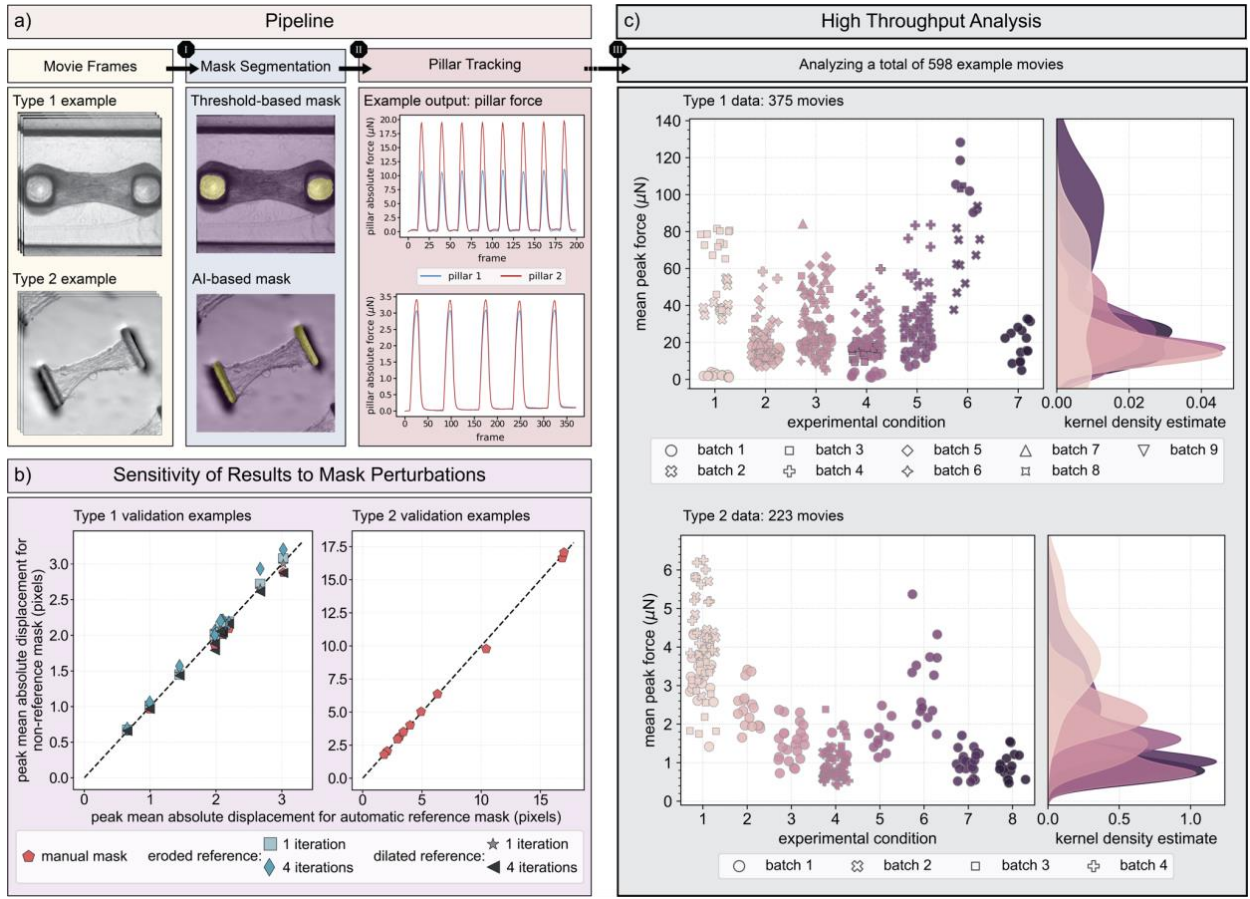

**Figure 1 – An overview of "MicroBundlePillarTrack," a Python-based package for the automated analysis of engineered cardiac microbundle contractility derived from tracked pillar or post deflection:** (a) Given movies of either "Type 1" or "Type 2" data as input (see "*Data*" Section under Methods for a description of these types), the software automatically segments the pillar regions, identifies marker points on the first valley frame for which the tissue is in a fully relaxed state, and tracks those points across all consecutive frames. From the tracked pillar displacements, the pillar force is straightforward to calculate by treating the pillars as beams with known elastic modulus and geometry. (b) To assess the quality of the automatically generated masks, we compare the tracked peak mean absolute displacement when an automatic mask (reference mask) is generated to when a manually generated mask (non-reference mask) is used for 11 typical examples of both data types. We also perform a sensitivity analysis of the tracked displacements in response to mask perturbations for "Type 1" data where we implement eroded and dilated variations of the automatic mask by a 5x5 kernel applied once or repeated 4 times. Overall, all displacement errors with manual masks lie below 6.2% for both data types and are less than 10% for the perturbed masks of "Type 1" data. (c) With "MicroBundlePillarTrack," it is straightforward to analyze large batches of data with no parameter tuning. For analysis, the required user input is limited to data type, frame rate (fps), length scale (μm/pixel), pillar stiffness (μN/μm), and tissue thickness (μm). Here, we summarize one possible output (mean peak force (μN)) obtained from implementing the software on: 1) 375 movies of "Type 1" data

with cardiac microbundles grown under 7 different experimental conditions and paced at 1 Hz (except for the data points marked with a black edge color which are beating spontaneously) run for the input parameters: frame rate of 30 fps, length scale of 4 µm/pixel, pillar stiffness of 2.677 µN/µm, and tissue thickness of 350 µm, and 2) 223 movies of "Type 2" data where 8 different experimental conditions were used to grow the spontaneously beating microbundles run for the input parameters: frame rate of 65 fps, length scale of 0.908 µm/pixel, pillar stiffness as specified for each condition in the "*Data*" Section, and tissue thickness of 10 µm. Such high-throughput analysis enables meaningful comparisons between different experimental conditions for the same platform, across different batches prepared under the same experimental conditions, and across different experimental platforms.

**Description**

The development of human induced pluripotent stem cells (hiPSCs) differentiated into cardiomyocytes (CMs) in Yamanaka's lab (Takahashi et al., 2007) more than a decade ago constituted a major breakthrough in the fields of basic and translational cardiovascular research. hiPSC-CMs' ability to self-assemble into beating cardiac tissues or more complex microscale cardiac tissue bundles (microbundles) allows for their employment in *in vitro* platforms for different applications including disease modeling, drug discovery, and regenerative tissue engineering (Brandão et al., 2017; Nakao et al., 2020; Hnatiuk et al., 2021). Despite the attractive opportunities that these engineered cardiomyocytes offer for advancing cardiac research, there is currently a major limitation: they resemble fetal CMs and are morphologically and functionally different from the cells in a mature cardiac tissue (Batalov et al., 2021). Different tissue culture platform designs that allow for mechanical, electrical, or magnetic actuation (Boudou et al., 2012, Xu et al., 2015, Ruan et al., 2016, Ronaldson-Bouchard et al., 2018; Javor et al., 2020; DePalma et al., 2021; Jayne et al., 2021) have been implemented to promote the maturity of hiPSC-CMs. In these platforms, hiPSC-CMs are often seeded between two flexible elastomeric pillars that allow for noninvasive actuation and quantification of the tissue function through pillar deflection measurements. However, this variability in maturation techniques gives rise to another challenge: performing reliable and reproducible quantitative comparisons of the functional behavior of microbundles grown across different pillar-based experimental setups.

The advancement in microscopic imaging techniques (Balasubramanian et al., 2023), especially in light microscopy, has enabled the collection of time-lapse movies of beating microbundles regardless of the experimental conditions and techniques used to create them. Having a uniform and consistent method for image-based data collection provides ample opportunities for developing tools for meaningful and reproducible data analysis across platforms. Currently, there exists a number of software tools to extract contractility metrics based on pillar deflection from this type of image-based data (Hansen et al., 2010; Oyunbaatar et al., 2016; Thavandiran et al., 2020; Dostanić et al., 2020; Javor et al., 2020; Tamargo et al., 2021; Rivera-Arbeláez et al., 2022; Méry et al., 2023; Tani et al., 2023). However, the majority of these computational tools are custom and are not broadly disseminated under open-source licenses for the benefit of the broader research community. And, currently available tools typically lack automation (with some notable exceptions such as the software developed by Rivera-Arbeláez et al.), and may not be readily extended to function on new datasets. Most importantly, these analysis tools do not share a unified definition of pillar deflection. Particularly, Tani et al. (2023), Rivera-Arbeláez et al. (2022), Tamargo et al. (2021), and Hansen et al. (2010), consider the distance between the actuated pillars to be the deflection, whereas Méry et al. (2023), Javor et al. (2020), Dostanić et al. (2020), and Oyunbaatar et al. (2016), compute the deflection of each pillar from its rest position and take the average of the two. Critically, the former approach will report both contractile forces and tissue stresses twice as high as the latter approach. We include, in the Extended Data Section, a schematic that further explains the difference between these two approaches.

Given this context, and in an effort towards establishing standardized quantitative metrics to assess the functional maturity of hiPSC-derived cardiac microbundles, we present here "MicroBundlePillarTrack," a versatile software for the automated segmentation, tracking, and analysis of pillar deflection. This software has been developed in Python and tested (to date) on two different tissue culture platforms. Our computational framework is built upon our prior work, "MicroBundleCompute" (Kobeissi et al., 2024), an optical flow-based tracking and analysis software of whole tissue deformation in cardiac microbundles. We have extensively validated our tracking pipeline against realistic synthetic data based on brightfield movies and against manual tracking (Kobeissi et al., 2024) and are confident that, for displacement magnitudes that exceed a single pixel, tracking errors due to imaging artifacts and noise fall below 10%. We have also performed additional validation to test the reliability of the automated mask segmentation step, where we perform a pillar mask sensitivity analysis (Figure 1b) to compare the peak mean absolute displacement obtained for an automatically generated mask (reference mask; Figure 1a) and a manually traced mask (non-reference mask; not represented here but appears visually similar to a reference mask with slight variation in size and shape) for 11 typical examples of two different data types as described under "*Data*" in the Methods Section (see Figure 1a for an example of each data type). We also perform a sensitivity analysis of the tracked displacements in response to mask perturbations (non-reference mask) for "Type 1" data only, where we implement eroded and dilated variations of the automatic mask by a 5x5 kernel applied once or repeated 4 times. Overall, all displacement errors with manual masks lie below 6.2% for both data types and are less than 10% for the perturbed masks of "Type 1" data.

With its fully automated workflow, "MicroBundlePillarTrack" facilitates the analysis of large batches of image-based data of beating cardiac microbundles without compromising the accuracy of the resulting contractility metrics. As a demonstration, we show that it is straightforward to analyze a total of 598 movie examples (Figure 1c) and perform relevant functional comparisons of microbundles grown with the same tissue culture platform but under different experimental conditions, and between microbundles created with different platform designs. Additionally, the software can be easily extended to accommodate data collected on new experimental setups. If the pillar designs fall out of the scope of our two data types, we include a code, on the software's GitHub repository, to adapt the automatic mask segmentation step to this new data. For the subsequent tracking and analysis steps, the same pipeline should work on new data types as long as the frame rate and the length scale of the brightfield movies are reasonable such that tracked displacements are not large between consecutive frames and the pillars exhibit an identifiable pixel texture.

Finally, we share "MicroBundlePillarTrack" under an open-source license and make a dataset of 1,540 brightfield example movies publicly available to encourage collaborative efforts within the tissue engineering research community and across different disciplines towards enhancing hiPSC-based technologies. We anticipate that the availability of large image-based databases of engineered cardiac microbundles would not only allow for the development of more robust software analysis tools, but would also promote other research endeavors including the design of more detailed and more accurate multiscale mathematical and computational models (Jilberto et al., 2023).

## Methods
### *Data*
We test "MicroBundlePillarTrack" on two different types of data as categorized in our previous work (Kobeissi et al., 2024) where "Type 1" data include movies of cardiac microbundles grown on standard experimental microbundle strain gauge devices (Zhang et al., 2021), whereas for "Type 2" data, non-standard platforms referred to as FibroTUGs (DePalma et al., 2023) are used.

Briefly, standard devices of "Type 1" consist of 6 wells, with each well containing 2 pillars with rectangular cross sections and spherical caps. The poly(dimethylsiloxane) (PDMS, Dow Silicones Corporation, Midland, MI) microbundle devices are cast from 3D printed molds and treated with 0.01% poly-l-lysine (ScienCell) followed by 0.1% glutaraldehyde (EMS) up until 3 days before seeding. This treatment promotes cell attachment to the pillar caps and the growth of the microbundle between the two pillars. Before seeding, the devices are sterilized with 70% ethanol and ultraviolet light and then incubated with a small volume of 2% Pluronic F-127 (Sigma) to prevent cell attachment to the bottom of the wells. Next, hiPSC-CMs, differentiated and purified with a protocol (Zhang et al., 2021) adapted from Lian et al. (2013), were seeded with human ventricular cardiac fibroblasts in a Matrigel (Corning) and fibrin (Sigma) extracellular matrix (ECM) solution with the growth medium replaced every other day. At days 5–15 after seeding, time-lapse videos of tissue contractions were taken at 30 Hz using a Nikon Eclipse Ti (Nikon Instruments Inc.) microscope with a 4x objective and an Evolve EMCCD camera (Photometrics), while maintaining a temperature of 37°C and 5% $CO_2$.

While the pillar stiffness is kept constant for this type of data at 2.677 µN/µm, the experimental conditions can be varied depending on the purpose of the study. Specifically, the 7 conditions plotted here in Figure 1c refer to microbundles with: 1) PGP1 hiPSC-CMs with an endogenous green fluorescent protein (GFP) tag on sarcomere protein titin (wild-type titin-GFP) and ventricular cardiac fibroblasts (vCFs) cultured in 150 µg/mL L-ascorbic acid 2-phosphate sesquimagnesium salt hydrate (ascorbic acid) and 0.01% dimethyl sulfoxide (DMSO) taken at Day 6, 2) wild-type titin-GFP hiPSC-CMs and vCFs cultured in 150 µg/mL ascorbic acid and 0.01% DMSO taken at Day 7, 3) wild-type hiPSC-CMs and vCFs cultured in 150 µg/mL ascorbic acid taken at Day 7, 4) PGP1 hiPSC-CMs and vCFs cultured in 150 µg/mL ascorbic acid taken at Day 7, 5) wild-type titin-GFP hiPSC-CMs and vCFs taken at Day 7, 6) wild-type titin-GFP hiPSC-CMs and vCFs that were pre-treated with 10 µg/mL mitomycin-C for two hours prior to microbundle seeding, inhibiting proliferation of vCFs, taken at Day 7, 7) wild-type titin-GFP hiPSC-CMs and vCFs that were pre-treated with 10 µg/mL mitomycin-C for two hours prior to microbundle seeding and cultured in 150 µg/mL ascorbic acid taken at Day 7. A complete summary of the experimental metadata is included with the published dataset (https://doi.org/doi:10.5061/dryad.sqv9s4nbg).

The FibroTUG platforms ("Type 2" data) consist of an array of fiber matrices generated by selective photo-crosslinking of electrospun dextran vinyl sulfone (DVS) (Davidson et al., 2020) suspended between a pair of PDMS cantilevers fabricated by soft lithography. The fiber matrices are then functionalized with cell adhesive cRGD peptides before differentiated and purified iPSC-CMs (DePalma et al., 2021) are patterned onto the matrices using microfabricated seeding masks cast from 3D-printed molds. After 3–21 days of seeding, time-lapse videos of the microbundle's spontaneous contractions are acquired at ~65 Hz on Zeiss LSM800 equipped with an Axiocam 503 camera while maintaining a temperature of 37°C and 5% $CO_2$.

For this "Type 2" platform, the stiffnesses of both the fiber matrix and the pillars can be adjusted by varying the photoinitiator concentrations during matrix crosslinking and cantilever height, respectively. The matrix alignment, on the other hand, can be controlled by changing the translation speed of the collection mandrel during fiber deposition. For the specific examples included here in Figure 1c, the 8 different experimental conditions correspond to: 1) matrix stiffness of 0.68 kPa, aligned fibers, and post stiffness of 0.41 µN/µm, 2) matrix stiffness of 0.68 kPa, aligned fibers, and post stiffness of 1.2 µN/µm, 3) matrix stiffness of 10.1 kPa, aligned fibers, and post stiffness of 0.41 µN/µm, 4) matrix stiffness of 17.4 kPa, aligned fibers, and post stiffness of 0.41 µN/µm, 5) matrix stiffness of 0.68 kPa, random fiber orientation, and post stiffness of 0.41 µN/µm, 6) matrix stiffness of 0.68 kPa, random fiber orientation, and post stiffness of 1.2 µN/µm, 7) matrix stiffness of 10.1 kPa, random fiber orientation, and post stiffness of 0.41 µN/µm, 8) matrix stiffness of 17.4 kPa, random fiber orientation, and post stiffness of 0.41 µN/µm. A more detailed summary of the experimental metadata is included with the published dataset (https://doi.org/10.5061/dryad.3r2280gqd).

*Code*

As mentioned previously, we develop "MicroBundlePillarTrack" as an adaptation of our previous software, "MicroBundleCompute" (Kobeissi et al., 2024). Specifically, this new extension is tailored to tracking pillar deflection in two different types of pillar-based microbundle platforms as described in the "*Data*" Section above.

The code is organized in three Python files: `create_pillar_mask`, `image_analysis`, and `pillar_analysis`. The first file contains all the functionalities for automated mask segmentation, while the remaining two scripts contain the bulk of the tracking functionalities. Additionally, we provide two user-friendly scripts, `run_code_pillar` and `run_code_pillar_batch`, that run the whole tracking pipeline and enable software users to perform the analysis, in single or batch mode, without having to go through the inner workings of the code. We summarize below the essential functions of the computational workflow. More details about the overall tracking pipeline, specific functions and outputs, tracking validation, and software installation and usage can be found on the GitHub repository (https://github.com/HibaKob/MicroBundlePillarTrack) and in Kobeissi et al. (2024).

Given an image sequence of the beating microbundle stored as consecutive individual frames (Figure 1a), the software, first segments the pillar regions in the first frame using either of the two provided approaches within `create_pillar_mask`: 1) a straightforward threshold-based segmentation that implements the local Otsu thresholding provided within the filters module in scikit-image 0.19.3 Python library (van der Walt et al., 2014), or 2) an AI-based approach that implements a fine-tuned version of the Segment Anything Model (SAM) (Kirillov et al., 2023). In case automatic segmentation fails, software users have the option to provide manually or externally generated binary masks where each pillar mask is a two-dimensional array in which the pillar domain is denoted by "1" and the background domain is denoted by "0". Critically, automatic segmentation might fail for 2 main reasons: 1) the input example falls outside of "Type 1" and "Type 2" data, or 2) the input example contains detrimental noise, has out-of-focus or obscured pillar regions, or is very dark with little variation in pixel intensities. In the case of the latter, we recommend that the user examines the tracked outputs generated for an external mask with extra care. For the former case, the user can fine-tune SAM on their own data by implementing the script that we have provided on the GitHub repository.

Following the segmentation of the regions of interest, the software tracks the position of these pillar regions across the frames following OpenCV's (Bradski and Kaehler 2008) pyramidal implementation of the Lucas-Kanade sparse optical flow algorithm (Bouguet et al., 2001). Fiducial markers identified as Shi-Tomasi "good features to track" (Shi et al., 1994) corner points are computed within the identified pillar masks on the first frame of the movie and tracked across all movie frames. We briefly note that, as with "MicroBundleCompute," parameters to tune OpenCV's `goodFeaturesToTrack` and `calcOpticalFlowPyrLK` functions are automatically adjusted to suit the specific input example.

From this initial tracking, the software checks whether or not the microbundle is in a fully relaxed state at the beginning of the movie (i.e. if the movie starts from a valley frame), identifies the first valley frame if it is not the first one, adjusts the movie to begin from this identified frame, and performs a second tracking of the adjusted movie if necessary. From the raw tracking outputs, that is, the column (horizontal) and row (vertical) positions of the marker points at each frame, it is straightforward to compute the pillars' mean directional and absolute displacements. Additional derived outputs include pillar twitch forces and tissue stress, as well as temporal outputs such as pillar contraction and relaxation velocities, full widths (or duration) at half maximum (FWHM), and full widths (or duration) at 80 maximum (FW80M). Specifically, pillar force is approximated by applying Hooke's law (Eq. 1) and using a cantilever beam equation (Eq.2) to obtain the pillar stiffness if the latter has not been determined experimentally (Legant et al., 2009; Das et al., 2022):

$$F = k\delta \qquad\qquad \text{Eq. 1}$$

where "$\delta$" is the pillar deflection or equivalently, the mean tracked pillar displacement (represented by $\Delta x_1$ and $\Delta x_2$ in the "Extended Data" figure) and "k" is the pillar stiffness as provided experimentally or derived using the equation:

$$k = \frac{6EI}{a^2(3L-a)} \qquad\qquad \text{Eq. 2}$$

where "$E$" is the material's elastic modulus (usually PDMS), "$a$" is the location of force application, "$L$" is the cantilever length, and finally "$I$" is the moment of inertia defined as a function of the cantilever's geometry. For beams with rectangular cross section, $I = \frac{wt^3}{12}$ where "$w$" is the pillar width and "$t$" is the pillar thickness, while for beams with circular cross section, $I = \frac{\pi D^4}{64}$ where "$D$" is the cylindrical pillar diameter. Mean tissue stress is then computed by dividing the pillar force by the tissue cross sectional area obtained from the tissue width as automatically measured by the software on the first valley frame, and tissue depth as approximated using three-dimensional imaging modalities. Finally, we note that all main software outputs described here are saved as ".txt" files and visualized as time series plots.

Additional functionalities of the software include detection and correction of tracked feature drift observed over the duration of the movie and the detection of irregular beats. Observed output drifts from 0 when in a fully relaxed state can be mainly attributed to excessive imaging noise and/or out-of-plane motion. To correct for this shift, we perform temporal segmentation whereby individual beats are identified from consecutive valley frames determined during preliminary tracking and split accordingly. All outputs are then computed per beat, taking the fiducial marker positions in the first frame of each beat as the baseline instead of the marker positions in the first frame of the entire movie. We note that this split is not performed automatically; rather, the user is warned that drift is detected and can choose to perform this temporal segmentation by setting the `split` parameter within the `run_code_pillar` or `run_code_pillar_batch` scripts to `True`. To execute this correction successfully, the code requires that the input movie spans at least 3 complete beats. In general, a minimum of 2 complete beats should be present in any movie to be analyzed. We mandate this requirement to enable more accurate outputs. As for irregular beats, a warning is issued to inform the user of this detection, but in the software's current state, no further action or analysis is performed.

**Reagents**

| REAGENT | AVAILABLE FROM |
| --- | --- |
| poly(dimethylsiloxane) (PDMS) | Dow Silicones Corporation |
| poly-l-lysine | ScienCell Research Laboratories |
| 0.1% glutaraldehyde | Electron Microscopy Sciences |
| 70% Ethanol | Multiple vendors |
| Pluronic F-127 | Sigma-Aldrich |
| Matrigel | Corning |
| human fibrinogen | Sigma-Aldrich |
| thrombin | Sigma |
| L-ascorbic acid 2-phosphate sesquimagnesium salt hydrate (ascorbic acid) | Sigma |
| dimethyl sulfoxide (DMSO) | Fisher |
| mitomycin C | EMD Millipore 475820 |
| dextran vinyl sulfone (DVS) | Synthesized in house as described in Davidson et al. (2020) |
| cRGD peptides ([Arg-Gly-Asp-D-Phe-Lys(Cys)]) | Peptides International |
| 0.6% (w/v) lithium phenyl-2,4,6-trimethylbenzoylphosphinate (LAP) | Colorado Photopolymer Solutions, Boulder, CO |

| CELL TYPE | DESCRIPTION, AVAILABLE FROM |
|---|---|
| human ventricular cardiac fibroblasts | Lonza (Cat. CC-2904), Cell Applications |
| wild-type titin-GFP hiPSC line | Synthesized as described in Sharma et al. (2018) |
| PGP1 hiPSC line | Coriell Institute (GM23338) |

**Extended Data**

Description: Link to "MicroBundlePillarTrack" GitHub repository
Resource Type: Software. URL:
https://github.com/HibaKob/MicroBundlePillarTrack/releases/tag/v1.0.0

Description: Links to the microbundle datasets
Resource Type: Dataset. DOI: https://doi.org/doi:10.5061/dryad.sqv9s4nbg ,
https://doi.org/10.5061/dryad.3r2280gqd

Description: Explanation for tissue and pillar force calculations
Resource Type: Image.

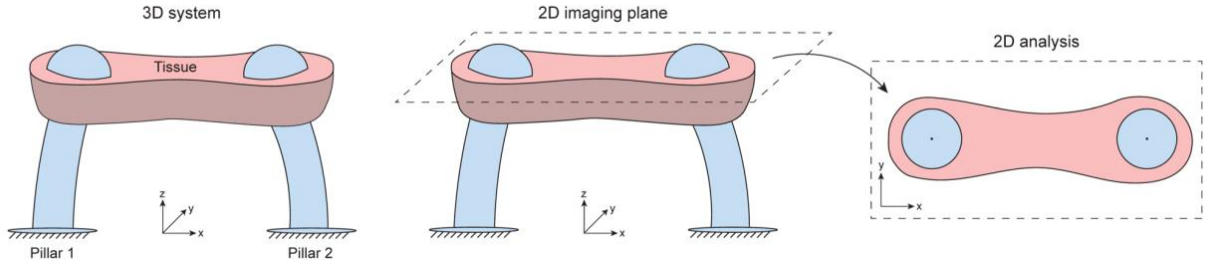

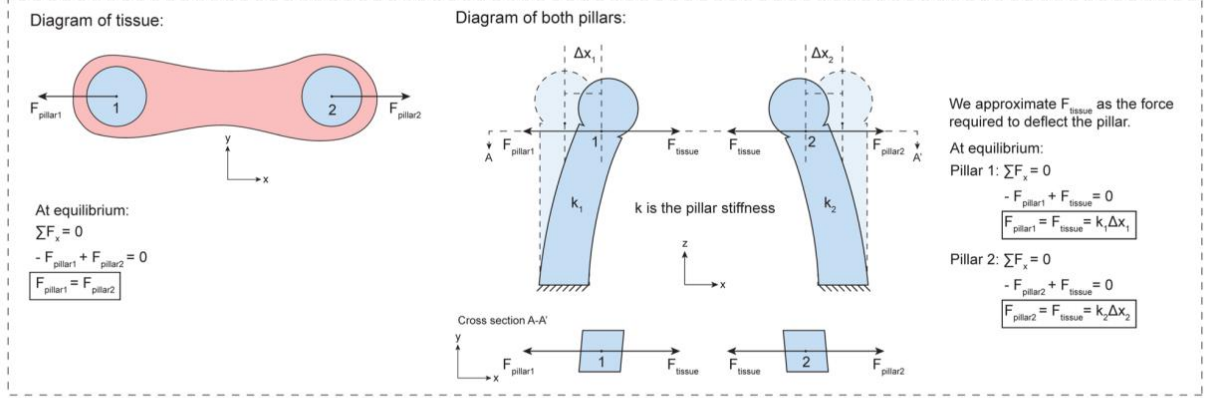

## Acknowledgements

We gratefully acknowledge the collaborative opportunities facilitated by the CELL-MET Engineering Research Center. We would also like to acknowledge the staff at Boston University libraries for providing advice regarding data dissemination practices.

## Funding Statement

This work was supported by the CELL-MET Engineering Research Center National Science Foundation ERC ECC-1647837. EL acknowledges support from the American Heart Association Career Development Award 856354, XG acknowledges support from the National Science Foundation Graduate Research Fellowship, BMB and SJD acknowledge support from NSF Award 2033654, SJD acknowledges support from NIH T32-DE007057 and NIH T32-HL125242, and JKE acknowledges support from NIH F31 HL158195-03. The funders had no role in study design, data collection and analysis, decision to publish, or preparation of the manuscript.